# TabVec: Table Vectors for Classification of Web Tables

Majid Ghasemi-Gol
University of Southern California
ghasemig@usc.edu

Pedro Szekely
Information Science Institute
szekely@isi.edu

## ABSTRACT

There are hundreds of millions of tables in Web pages that contain useful information for many applications. Leveraging data within these tables is difficult because of the wide variety of structures, formats and data encoded in these tables. TabVec is an unsupervised method to embed tables into a vector space to support classification of tables into categories (entity, relational, matrix, list, and non-data) with minimal user intervention. TabVec deploys syntax and semantics of table cells, and embeds the structure of tables in a table vector space. This enables superior classification of tables even in the absence of domain annotations. Our evaluations in four real world domains show that TabVec improves classification accuracy by more than 20% compared to three state of the art systems, and that those systems require significant in domain training to achieve good results.

## 1 INTRODUCTION

There are hundreds of millions of tables in web pages. Web tables present the data in a semi-structured way which makes them useful for data integration tasks, e.g. *schema auto-completion*, *query answering*, and *knowledge base (KB) completion* [3]. However, resolving the semantics of table cells, identifying table headers and mapping them to schema, and extracting relations from tables is challenging in three ways: 1) Only about 2% of tables on the web are data tables (contain useful information) [3], while the rest are non-data tables. Leveraging the useful data in the web-tables is dependent on finding useful tables first. 2) A data table can appear in different formats some of which are shown in Fig. 1. 3) The web-tables often provide little context, i.e. there is little or no supporting text surrounding the table and table cells contain short phrases, words, numbers, or just abbreviations. Despite efforts have been made to target these challenges , state of the art systems rely heavily on expert knowledge. More specifically, these systems usually require a domain ontology or rich KB to extract reliable relations from web tables [9, 18], or lots of labeled data for training a machine learning model [5, 8].

In this paper, we target the problem of discovering data tables and determining how data is organized within them, on a large corpus of web tables. We propose *TabVec*, a novel approach to embed tables in a vector space. We first train a word vector space model on the tables from the target domain, using our definition of word context, specific to tables. We then introduce a method to calculate table vectors deploying the word vector space, in order to embed structural features of tables. These steps are done in an unsupervised manner, given the raw web pages from a specific domain. The resulting table vector space enables clustering of tables according to their formats, within a specific domain. As a use case of this clustering, we use the web table taxonomy introduced by Crestan et. al [8] to classify the tables by type. We focus on five dominant table types: *relational*, *entity*, *matrix*, *list*, and *non-data*, examples of which are shown in Fig. 1. Although our method can be applied to tables from other kinds of online documents (CSV sheets, PDF documents, etc.) we focus on web tables in this paper, since they are easily accessible and easy to parse.

Existing methods for table type classification rely mostly on existing KBs [15], require lots of annotations [19], and are focused on generic domains [11], such as the *Common Crawl data*. In this work, we explore less investigated domains for which no existing KBs exist (securities fraud, illegal firearm sales, and human trafficking). Furthermore, for such domains, providing annotations is cumbersome and methods that require little or no annotation are helpful. Our method clusters the tables within a specific domain in an unsupervised manner, which consequently enables the user to examine the clusters and determine the table type for each cluster with minimal effort. Moreover, our proposed method can easily incorporate training data to enhance the performance of table type classification, i.e. the table vectors can be treated as a feature space for supervised classification of tables.

We study the web tables from four domains, three of which are different from the generic domains investigated before in three aspects, 1) there exists no KB or schema for them, 2) they are long tailed, i.e. the web pages in these domains look very different, and 3) they have noisy information, e.g. they contain incomplete, erroneous, or obfuscated data. We call these *unusual domains*. We also include a generic domain from Web Common Crawl data. Our Experiments show that, in absence of domain annotations, *TabVec* achieves on average 0.40 higher F1 micro score compared to state of the art systems, over the four domains we study.

The contributions of this paper are:

(1) introduce multiple definitions for word context in a table and explore how these definitions affect the quality of word vector space (and consequently the quality of table vectors and table clusters),

(2) embed semantic and syntactic features of a table cell in a vector representation, and introduce an effective and efficient method for table vector calculation deploying these cell vectors, in order to embed structural features of tables,

(3) employ table vectors to cluster tables according to their type, and evaluate the quality of resulting clusters on four real-world datasets.

The rest of this paper is organized as follows. Section 2 presents our approach to table type classification. Section 3 discusses related work. Section 4 presents our experimental setup and results, and section 5 presents conclusions and directions for future work.

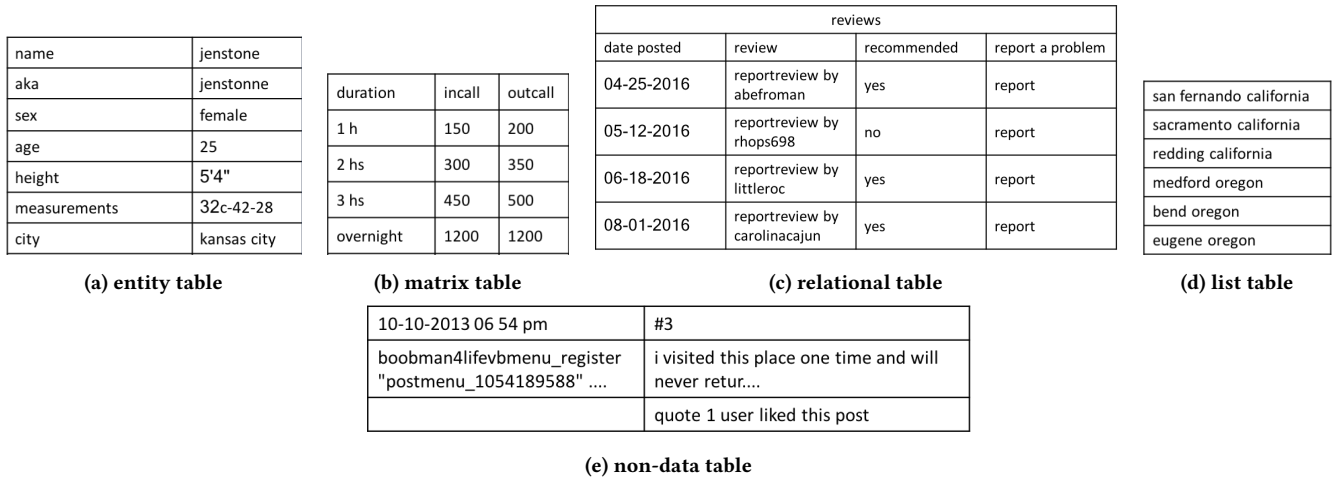

(a) entity table    (b) matrix table    (c) relational table    (d) list table

(e) non-data table

**Figure 1: Examples of table types from human trafficking advertisements.**

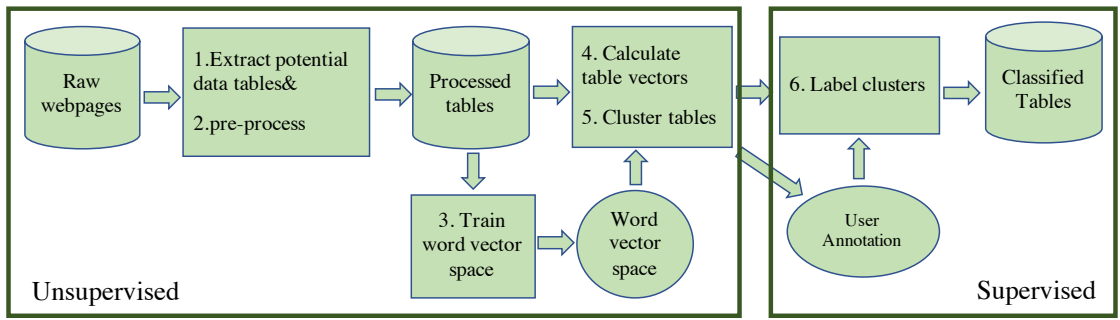

**Figure 2: The overall view of TabVec.**

## 2 METHOD

Word vector models have proved useful for learning the semantics of words given the context they appear within the corpus [17]. In our approach, we create a vector space model for web tables to represent the semantic and syntactic structure of web tables in a specific domain, and use these vector representations to identify the table type. The overall view of our system is shown in Fig. 2. Given a set of raw web pages from a specific domain, TabVec will 1) extract all the web tables within the web pages and prune out obvious non-data tables, 2) pre-process the tables to remove all the unnecessary tokens within the table cells (styles, HTML tags, etc.), and regularize some special tokens (e.g. dates and numbers), 3) construct a word vector space model (word vectors) based on the pre-processed web tables, 4) using this model, calculate a vector for each web table, 5) use the table vectors to identify clusters of table types. Up to here, all the steps are done in an unsupervised manner. In step 6) the user then manually labels the clusters to assign a proper table type to each cluster (possibly the majority of table types within the cluster). The remaining of this section discusses the steps of our approach in more details.

### 2.1 Table Detection and Extraction

We build a system that given the raw HTML webpage, extracts potential data tables. Since only a small portion (less than 2%) of tables in webpages on the Internet are data tables [3], this is an important step to prune non-data tables from further processing. Our approach, based on Cafarella's work [3], extracts HTML tables with no inner tables, and removes tables with fewer than 3 columns and fewer than 3 rows. Our domains contain many $2 \times 2$ data tables so our approach preserves all tables with at least 2 columns or 2 rows.

### 2.2 Table Pre-processing

Our studies revealed that formatting information within table cells is heterogeneous and adds no significant information regarding the table type. Consequently, TabVec removes HTML tags and styles (except <th>, <td>, <href>, and  tags, which proved useful in previous work) and keeps the cell content shown in the rendered webpage in a browser. The cell content is stored lowercased, and the metadata about cell formatting is encoded by appending to the text the keywords *TH*, *TD*, *HREF*, and *IMG* corresponding to tags <th>, <td>, <href>, and  present in the cell. To produce an $N * M$ array representing the table content, *colspan* or *rowspan* are unrolled, copying the content into individual cells. This unrolling simplifies the calculation of word context vectors. In domains where a semantic typer exists to tag cell contents with a semantic label (e.g., AGE), the semantic label is also appended to the corresponding cell content. This is shown in the example in Fig. 3.

For data types which have numbers in their range (e.g. price, age, date, etc.), the exact tokens are very sparse and we may not be able to train a good word vector space. Therefore, regularizing these values may improve the quality of word vector space. We use a very simple way to regularize numbers, where digits are regularized by substituting a special character for each digit. We substitute the digits with $X$ in the example shown in Fig. 3. This method categorizes the numbers by the number of their digits (e.g. dates will look like $XX/XX/XXXX$). This is not the universal way to regularize numbers, and more complex regularization methods may be used, however, this simple approach proves to improve the performance of TabVec in our experiments.

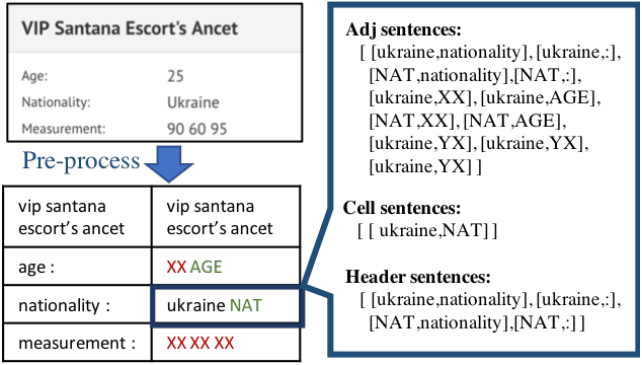

**Figure 3: an example of table pre-processing and different word context definitions.**

## 2.3 Constructing The Word Vector Space

When constructing a word vector model, we need to define a context for every word mentioned in the document. In the conventional use cases of word embeddings, human language sentences are used to learn the context of the words. There are two main issues regarding this choice in our domains, 1) the webpages sometimes contain no text out of the table, and when they contain text, it is often very short, 2) many words are exclusive to the tables, and do not appear in the text. So we need to redefine context for the words within the tables. We define three additional contexts the words in data tables, 1) the text within each table cell, 2) the text in the corresponding header row/column, and 3) the text in adjacent cells (top, bottom, left, and right). Together with the text surrounding the table in the HTML page, we have four different context definitions for the words in the corpus.

For each of these different contexts, TabVec generates context sentences, and use these context sentences to construct a word vector model. For the text surrounding the web table, these sentences are simply tokenized human language sentences. For the text within the table cell, the context sentence will be the tokenized cell text. The context sentences for header row/col, and adjacent cell, will be the cross product of the tokens within a cell and corresponding header cell, and adjacent cells respectively. More formally let us define an $N \times M$ table $T$ as an array of table cells:

$$T = \{ c_{ij} \; ; \; 1 \le i \le N, 1 \le j \le M \}$$

where each cell $c_{ij}$ is a collection of words:

$$c_{ij} = \{ t_{ij}^k \; ; \; 1 \le k \le K_{ij} \}$$

and $K_{ij}$ is the number of words in $c_{ij}$. The equations below show the formulas to construct cell sentences ($S_c$), header cell sentences ($S_h$), and adjacent cell sentences ($S_a$). In these equations $c_{ij} \times c_{i'j'}$ denotes the cross product of the tokens in $c_{ij}$ with the ones in $c_{i'j'}$. $S_{st}$, a collection of tokenized sentences denotes the context sentences from the text surrounding a web table. Fig. 3 shows an example of $S_c$, $S_h$, and $S_a$.

$$S_c = \{ c_{ij} \; ; \; 1 \le i \le N, 1 \le j \le M \} \quad (1)$$

$$S_h = \{ c_{ij} \times c_{i1} \; ; \; 1 \le i \le N, 2 \le j \le M \}$$
$$\cup \{ c_{ij} \times c_{1j} \; ; \; 2 \le i \le N, 1 \le j \le M \} \quad (2)$$

$$S_a = \{ c_{ij} \times c_{i,j+p} \; ; \; 1 \le i \le N, 2 \le j \le M \}$$
$$\cup \{ c_{ij} \times c_{i+p,j} \; ; \; 1 \le i \le N, 2 \le j \le M \} \quad (3)$$
$$\text{where:} \quad -w \le p \le w, \; p \ne 0$$

We use *random indexing* [14] to construct a word vector space in this paper. In this algorithm, given a vector dimension $d$ and a window size $w$, first base vectors are build for the words in the corpus (we prune out low and high frequency words). The base vectors are created by putting -1 and 1 in random positions in the vector (other positions are zero). We set number of these positions to be 4 (two 1's and two -1's). A word vector is constructed by summing up all the base vectors of other words that are within the $w$ distance of the word in all sentences (this is done incrementally by processing the sentences one by one and updating the word vector space accordingly; in our implementation, we calculate the word vectors as a MapReduce job). This makes this algorithm easy to implement, easy to parallelize, and very scalable .

## 2.4 Calculating Table Vectors

Every word in a table cell $t_{ij}^k$ can now be presented as a vector $v_{t_{ij}^k}$. So, let us redefine our cell notation as $\hat{c}_{ij} = \{ v_{t_{ij}^k} \; ; \; 1 \le k \le K_{ij} \}$. We expect that each cell in the table presents a single concept, and the words within the cell are related to that concept. We expect that the word vectors $v_{t_{ij}^k}$'s are close to each other, so a combination of these word vectors can be used to represent that concept. In order to enhance resilience to noise, we take the median of these vectors rather than the average (the median of the vectors is computed as the vector of medians of each entry). These cell vectors embed the semantic meaning of each table cell, and the distance between two cell vectors measures the similarity of the concepts associated with them.

A table type defines how different concepts are organized in a table, and the cell vectors can be used to analyze this organization and consequently identify the table type. More specifically, a table type defines how concepts are distributed in cells within table rows, table columns, and the whole table. For example, if we consider relational vs. matrix table type, the cells in every column contain similar concepts in both table types, while the cells in every row contains similar concepts only in matrix tables. We argue that analyzing semantic consistency for the cells in table rows, table columns, and the whole table is sufficient to identify the table type.

We quantify the similarity in semantic meaning over a set of cells by computing the *deviation from the mean* and *deviation from the median* for the cell vectors (common measures of variability and consistency of sample or population). Earlier, we used the median of word vectors to calculate a cell vector, however, here we use both deviation from median and mean to calculate consistency. This has two reasons, 1) for calculating the cell vectors we assumed the words in a cell are related to the same concept while here this may not be the case, 2) the deviation from the median is a resilient consistency measure, and deviation from the mean captures the noise. For example, considering the cells within a relational table column, deviation from the median is small while deviation from the mean is high, since the header cell has a very different vector compared to the data cells. Comparing this with the cells in a list table column, both deviation from the mean and the median are small.

Let us redefine our table notation by using the cell vectors:
$$\tilde{T} = \{\tilde{c}_{ij} \; ; \; 1 \leq i \leq N, 1 \leq j \leq M\}.$$
We denote deviation from median with $dev^{median}$ and deviation from mean with $dev^{mean}$. We calculate six vectors for each data table, $dev^{mean}_{rows}$, $dev^{median}_{rows}$, $dev^{mean}_{cols}$, $dev^{median}_{cols}$, $dev^{median}_{all}$, $dev^{median}_{all}$. We show how the vectors for deviation from the median are calculated in equations 4-6. The vectors for deviation from the mean are calculated similarly. In these formulas, element-wise power of 2 is used for calculating deviation from the mean/median vectors, and the resulting deviations are vectors. Also, note that $median_i(v_{c_{ij}})$, $median_j(v_{c_{ij}})$, and $median_{i,j}(v_{c_{ij}})$ denote the median of the cell vectors in row i, column j, and the whole table respectively.

$$dev^{median}_{rows}(\tilde{T}) = \frac{1}{N} \sum_i \left( v_{c_{ij}} - median_i(v_{c_{ij}}) \right)^2 \quad (4)$$

$$dev^{median}_{cols}(\tilde{T}) = \frac{1}{M} \sum_j \left( v_{c_{ij}} - median_j(v_{c_{ij}}) \right)^2 \quad (5)$$

$$dev^{median}_{all}(\tilde{T}) = \frac{1}{N \times M} \sum_{i,j} \left( v_{c_{ij}} - median_{i,j}(v_{c_{ij}}) \right)^2 \quad (6)$$

We define a table vector to be the concatenation of these vectors. Fig. 4 illustrates an example of the table vectors for different table types, reduced to 2D space using *t-SNE* [26]. As we can see, although not all the points are fully separated, there are distinguishable table type clusters detected by the table vectors.

## 2.5 Determining Table Type

The table vectors created in previous section, embed the structural features of tables. This enables forming clusters of tables with similar structure (table type in our problem). Number of optimum clusters may differ from one domain to another, e.g. due to lack of some table types in a domain, or multiple separate clusters emerging for the same table type due to different orientations. The best number of clusters can be found using different existing methods in the literature, and we use *silhouette score* [23] in this paper. Note that this is done without any need for user annotation, and can be efficiently done on a large set of tables from each domain. We use *Kmeans* [16] clustering algorithm in this paper.

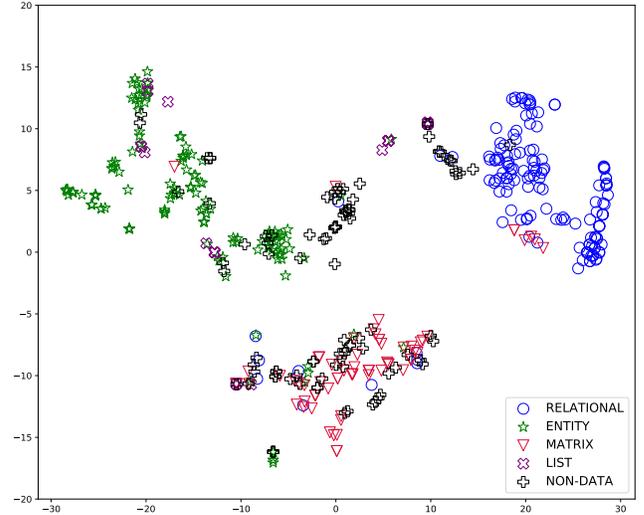

**Figure 4: An example of table vectors shown in 2D space.**

The clustering algorithm yields clusters of tables, but does not assign a label to each cluster to identify the table type. There are several automatic cluster labeling techniques used in information retrieval from documents, e.g. mutual information, title labels, centroid labels, etc. [7] In terms of table types, these auto-labeling techniques do not produce the desired result because the table type is not found within the web table itself. Human intervention is necessary to label the clusters. We collect the closest point to each cluster centroid, and ask users to label manually label the tables corresponding to these points. TabVec assigns the majority label to the corresponding cluster.

## 3 RELATED WORK

Semi-structured data sources (primarily web tables) have been of interest for for information retrieval and knowledge extraction from early days. Research from Google (Octopus system) [2–4] pioneered web-scale knowledge extraction from the tables. In their research they try to extract relations from the web tables. Their system combines search, extraction, data cleaning, and data integration. Web tables have been afterwards considered for relation extraction [28, 29], KB creation and enhancement [10, 18, 21], data augmentation [1, 22], and query answering [6, 20, 24, 25].

Although web tables have proved to be useful sources, detecting useful information from millions of tables on web is challenging. Researchers have been working on methods to detect useful data tables from web. Methods have been proposed to distinguish data tables [30]. Also, some research is focused on labeling of table cells

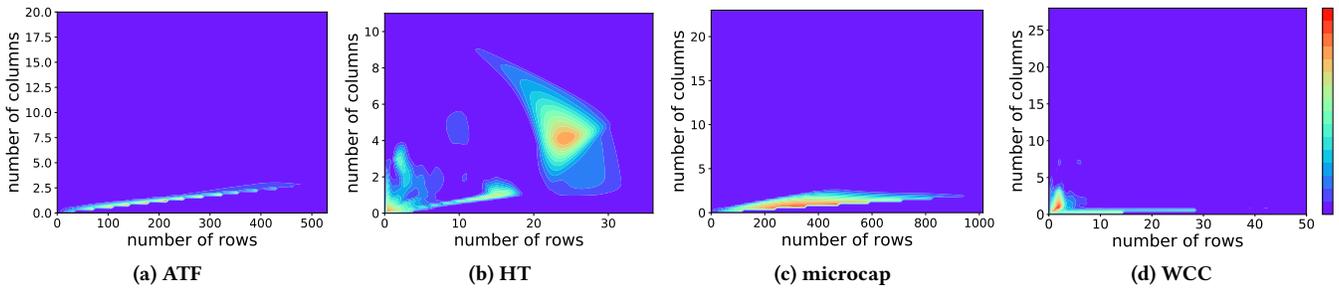

**Figure 5: number of rows and columns for web-tables in each domain. Note that range of the axises differ in different domains.**

(a) ATF  (b) HT  (c) microcap  (d) WCC

[9, 12] or table columns [27] to leverage web tables for different information integration tasks. These methods usually use an existing KB to infer the label of cells or columns.

Moreover, the fact that knowing how the data is organized in a table, significantly helps with making sense of the table cells, motivated a more fine grained classification of table types. Crestan et. al introduced seven different table types for web tables (listings, attribute/value, matrix, enumeration, form, navigational, and formatting), and presented a feature based classification technique, using structural features, and syntactic content features (e.g. number of empty cells, number of cells containing number, etc.), to identify the table type [8]. Later, Eberius et al. enhanced their model by introducing more features in [11], and used their system to build a table corpus of web tables from 3.6 billion web pages. Li et. al improved performance of previous systems by adding more features based on the semantic of table content for a case study of financial documents [15]. More recently, Nishida et. al proposed a supervised deep learning method (TabNet) using a hybrid deep neural network architecture, based on *Hierarchical attention networks* [19]. They evaluated their system on the common crawl dataset, and reported significant improvement compared to previous feature based methods. They used more than 60,000 annotated web tables, to train their model. Unlike these methods, TabVec does not require manually annotated training data, nor an existing knowledge base. We use user annotations for examining the clusters, but the effort of doing this task is minimal compared to previous methods. Moreover, word vector spaces are not used in prior work. Although, Nishida et. al mentioned that their system can leverage a word space model, the fact that they use a supervised approach to train their model differentiates their system from TabVec.

Deploying vectorization methods for web tables is rather unexplored and, to the best of our knowledge, there are few works in this area. Gentile et. al use table embeddings for blocking step in entity matching problem [13]. Their work is different from TabVec in that they do not determine the table type using the table vectors. They assume that the header rows, and attribute value relationship (for entity model) is known in order to create the context sentences for words within tables.

## 4 EXPERIMENTS

We evaluate TabVec [1] on four real-world datasets. Three of these datasets are from unusual domains, and one is a sample from Common Crawl. We use different configurations for our system, and compare the performance of our system with three state of the art systems for table type classification, which use supervised models. Note that our goal is to minimize user intervention; so, in our experiments, we use pre-trained models for our baselines when available. In absence of domain annotations, TabVec outperforms the baselines on all our experimented domains.

### 4.1 Experiment setup

*4.1.1 Datasets.* In our experiments we use raw web pages extracted from three unusual domains, *human trafficking advertisements (HT)*, *fire arms trading (ATF)*, and *microcap stock market (microcap)*, We selected these unusual domains because of the social good impact that accurate knowledge extraction can have by helping investigators identify fraud on the web. In addition, as Fig. 5 shows, these domains are technically interesting in the way that they contain unusually large or small tables. Also, we use a random sample of *July 2015 Common Crawl (WCC)* as a generic domain to compare our system with the state of the art systems. For our unusual domains, web pages are all crawled from websites and forums from the open web, however, due to privacy agreements the raw web pages cannot be released to public [2]. ATF domain is full of very small tables (description of the fire arms), HT domain contains both small tables (adult service description) and medium size tables (adult service listings), and microcap domain has small tables (e.g. description of the company) and medium and large size tables (e.g. stock value listings). The WCC domain consists of small and medium size tables. Table 1 shows the number of tables extracted from each domain as a heat-map plot.

*4.1.2 Constructing groundtruth.* We sample 500 tables from each domain, and manually collect tables of different types from these samples. We repeat the sampling until we have at least 80 tables for each type, and after each iteration if we already have more than 100

---

[1] TabVec can be accessed at: https://github.com/majidghgol/tabvec. The input to the system is the output of our extraction toolkit (ETK): https://github.com/usc-isi-i2/etk
[2] At the time of writing this paper we are working with our sponsor to release the corpus of tables extracted from the web pages after coding personal identifiable information. The webpages and groundtruth that we used from WCC domain, along with all the intermediate outputs from TabVec can be accessed at: https://www.dropbox.com/sh/tmy29tz9nzrcpo6/AAC356QkWE39Ho3HiYQHKcq5a?dl=0

|  | pages | tables | tables/page | tables $> 1 \times 1$ |
|---|---|---|---|---|
| ATF | 200K | 496K | 2.5 | 454K |
| HT | 60K | 30K | 0.5 | 24K |
| microcap | 200K | 337K | 1.7 | 288K |
| WCC | 9K | 24K | 2.6 | 20K |

**Table 1: number of pages and tables in each dataset.**

|  | entity | relational | matrix | list | non - data | sum |
|---|---|---|---|---|---|---|
| ATF | 94 | 114 | 81 | 0 | 106 | 395 |
| HT | 174 | 150 | 115 | 125 | 118 | 682 |
| microcap | 134 | 159 | 81 | 0 | 97 | 471 |
| WCC | 95 | 96 | 39 | 35 | 121 | 386 |

**Table 2: number of tables for each table type in the groundtruth for each dataset.**

tables for a type, we stop adding more of them to the groundtruth (to avoid biasing the groundtruth). We did not find any list tables in the *ATF*, and *microcap* domains, also for *WCC* domain we could not meet the 80 cap for matrix and list types. Table 2 shows a summary of our groundtruth for different domains.

*4.1.3 Baseline systems.* For our baseline systems, we choose three state of the art systems, *webcommons extraction framework*[3] (webcommons), *Dresden Web Table Corpus extractor*[4] (DWTC), and a neural network based model (TabNet) [19]. For *webcommons* and *DWTC*, we use pre-trained models that comes with them. For *TabNet*, we report the accumulative result of 10-fold cross-validation.

*4.1.4 Parameter selection.* We choose $w = 1$ for generating adjacent cell sentences. We only consider the cells containing a <th> tag as header cells when generating header cell sentences. Also, whenever we need cross product of two cells, we randomly sample 50 elements from the cross-product in case it results in more than 50 word pairs. In the random indexing method, we choose $w = 2$, and set four positions to -1 and +1 (two each) in the base vectors. Moreover, we use K-Means algorithm for the clustering task.

*4.1.5 Choosing number of clusters and vector size.* We calculate mean Silhouette Coefficient on all the table clusters varying the number of clusters to determine the optimal number of clusters for each of our domains. This gives a measure of inter- vs. intra-cluster distance for table vectors, and uses no annotations. Fig. 6a shows the Silhouette score for different domains, using 4, 6, 8, 10, 12, and 14 clusters. $n = 12$ shows high Silhouette score for all of our domains, so we select n to be 12 throughout our experiments.

Selection of vector dimension (d) in the random indexing algorithm affects the quality of word vector space, i.e. very small d results in higher collision probability in base vectors (and consequently table vectors and table clusters), and very large d makes the distance between semantically different words insignificant (many dimensions have a value of zero in many word vectors). We choose d by calculating the F1 micro score for our classification task varying d (groundtruth annotations are used). Fig. 6b shows the result of this experiment. As we would expect, for HT and WCC, where we have fewer number of tables (there would be fewer word co-occurrences and smaller word vectors), smaller d is favorable.

On the other hand, for ATF and microcap, which contain many more tables, larger d works better. In our datasets, selecting $d = 200$ results in good performance for all domains, and the rest of the experiments in this paper is done by choosing d to be 200.

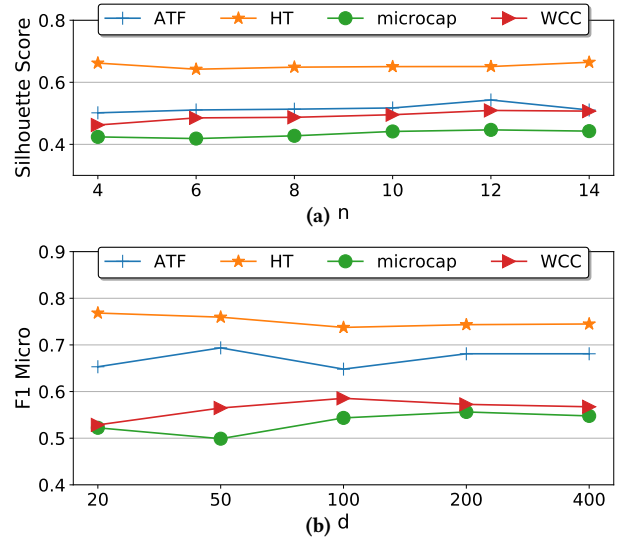

**Figure 6: (a) Silhouette score varying number of clusters ($d = 200$) (b) F1 micro score varying vector dimension ($n = 12$). For both figures context is table cell text, and regularization is performed.**

## 4.2 Experiment results

We studied the performance of our system with different sets of configurations on the different domains. We specifically studied how context sentence generation, and table preprocessing affects the performance of the system.

*4.2.1 Effect of context selection.* We study the effect of using different context sentences in constructing the word vector space model. Fig. 7a illustrates F1 micro score for different domains, and Fig. 7a shows the type specific (F1 score) for HT domain, using different definitions for context: 1) only surrounding text (T), 2) only cell text (C), 3) T+C, 4) C plus header cells (C+H) 5) C+H plus Adjacent cells (C+A+H), 6) T+C+A+H. As we expected, T sentences do not result in a good word vector space. C sentences show good performance for all the domains. H and A context sentences help with the performance of TabVec in WCC domain, but not for other domains. This may be due the fact that WCC domain contains better formatted web tables (e.g. using <th> tags only for headers). For the human trafficking domain, we can see that each of the context sentences performs better than others for specific table types.

*4.2.2 Effect of regularization.* We study the effect of preprocessing on classification performance by performing our experiments w/ and w/o word regularization. Fig. 8a depicts effect of regularization on F1 micro score for different domains, and Fig. 8b shows class-specific F1 scores for table type classification in HT domain. As we expected, token regularization results in better word vectors and improves the overall performance on all domains. We

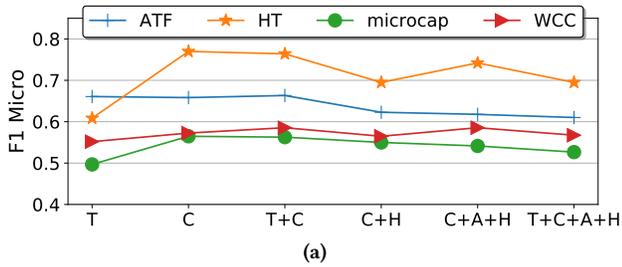

**(a)**

| context | perclass F1 score | | | | | F1-M |
|---|---|---|---|---|---|---|
| | R | E | M | L | ND | |
| T | 0.79 | 0.71 | 0.57 | 0.13 | **0.52** | 0.61 |
| C | 0.90 | **0.85** | 0.67 | 0.81 | 0.45 | **0.77** |
| T+C | **0.91** | 0.79 | 0.67 | 0.83 | 0.45 | 0.76 |
| C+H | 0.72 | 0.81 | 0.53 | 0.79 | 0.44 | 0.70 |
| C+A+H | 0.85 | 0.74 | **0.78** | 0.80 | 0.47 | 0.74 |
| T+C+A+H | 0.69 | 0.76 | 0.59 | **0.85** | 0.40 | 0.70 |

**(b)**

**Figure 7: Effect of context sentences. (a) classification accuracy varying context sentences for different domains. (b) F-1 scores by choosing different context sentences from the data to create the word embeddings (HT domain). In both figures** $n = 12$, $d = 200$ **and token regularization is performed.**

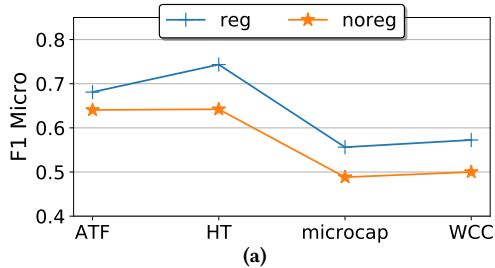

**(a)**

| | per class F1 score | | | | | F1-M |
|---|---|---|---|---|---|---|
| | R | E | M | L | ND | |
| w/ regularization | 0.90 | **0.85** | **0.67** | **0.81** | 0.45 | **0.77** |
| w/o regularization | **0.91** | 0.78 | 0.33 | 0.62 | **0.48** | 0.67 |

**(b)**

**Figure 8: Token regularization effect. (a) classification accuracy for different domains w/ and w/o regularizing tokens (reg vs. noreg) (b) F-1 score for the cases w/ and w/o token regularization in HT domain. In both figures** $n = 12$, $d = 200$ **context sentences is set to** *C*.

observe that only for relational and non-data tables, regularization marginally degrades the F1 score (it causes different tokens look the same, so in case of non data tables it may mistakenly regularize tokens from different concepts into the same form).

*4.2.3 Performance evaluation.* Based on the observations mentioned in previous sections, we decided to use *C* as word context sentences (C sentences showed to perform good on all the domains, and although one can decide which sentences to use for a specific domain, in this paper we would like to minimize user intervention), and regularize the tokens. Table 3 shows the class-specific and overall performance of the systems on different domains. TabVec outperforms other systems in each type of tables, and also in terms of F1 micro, for all the domains. Note that microcap and WCC domains contain many numeric tables, which hardens distinguishing different table types. Figure 9 depicts normalized confusion matrices for the classification task, for different systems on different domains. the X axis shows the predicted label and Y axis is the true label. Darker color means there are more data points assigned to a cell, and we expect the colors on the diagonal to be darker for a better classification performance. As we would expect, non-data tables are many times misclassified as data tables (mostly as entity and relational tables). This is expected since non-data tables contain textual information, and are usually medium size (e.g. a forum post). Also, the performance of TabVec for entity and matrix tables is lower. Matrix tables usually contain numeric data, and can be mistaken as relational tables (they have headers and similar values in each column). Entity tables contain attribute name and values for a single entity, where relationship of name and values are hard to embed (may need many occurrences), so they are sometimes classified as non-data tables.

| | system | per class F1 score | | | | | F1-M |
|---|---|---|---|---|---|---|---|
| | | R | E | M | L | ND | |
| ATF | webc | 0.22 | 0.06 | 0.03 | - | 0.05 | 0.09 |
| | DWTC | 0.20 | 0.16 | 0.03 | - | 0.04 | 0.11 |
| | TabNet | 0.24 | 0.12 | 0.34 | - | 0.18 | 0.25 |
| | TabVec | **0.65** | **0.57** | **0.68** | - | **0.73** | **0.66** |
| HT | webc | 0.46 | 0.38 | 0.00 | 0.00 | 0.07 | 0.32 |
| | DWTC | 0.35 | 0.42 | 0.00 | 0.00 | 0.25 | 0.32 |
| | TabNet | 0.06 | 0.40 | 0.03 | 0.21 | 0.18 | 0.26 |
| | TabVec | **0.90** | **0.85** | **0.67** | **0.81** | 0.45 | **0.77** |
| microcap | webc | 0.44 | 0.45 | 0.03 | - | 0.04 | 0.34 |
| | DWTC | 0.41 | 0.47 | 0.07 | - | 0.00 | 0.37 |
| | TabNet | 0.30 | 0.26 | 0.21 | - | 0.06 | 0.24 |
| | TabVec | **0.62** | **0.61** | **0.42** | - | **0.49** | **0.56** |
| WCC | webc | 0.27 | 0.13 | 0.09 | 0.00 | 0.11 | 0.15 |
| | DWTC | 0.22 | 0.31 | 0.00 | 0.00 | 0.06 | 0.17 |
| | TabNet | 0.09 | 0.41 | 0.09 | 0.11 | 0.25 | 0.27 |
| | TabVec | **0.67** | **0.41** | **0.22** | **0.48** | **0.69** | **0.57** |

**Table 3: Evaluation results. Note that in this table** *R, E, M, L, ND* **stand for relational, entity, matrix, list, and non-data table types. Also,** *F1-M* **stands for F1-micro score, and** *webc* **stands for webcommons.**

Pre-trained models of webcommons and DWTC are tuned for relational and non-data tables, which comply with our experiments, and these systems do poorly on matrix tables. Please note that these systems do not have list tables in their pre-trained models. TabNet shows to learn entity and list tables faster than other types, however, it seems that it needs more annotations to train a good model.

In order to explore whether domain annotations may improve performance of TabVec compared with baseline systems, we do a 10 fold cross-validation on our datasets, for TabVec and DWTC (DWTC showed similar behavior to webcommons in previous experiments, and we already did cross-validation for TabNet). We use *KNN* algorithm ($k = 5$) for TabVec (table vectors are treated as features), and *Random Forest* algorithm ($n = 100$) for DWTC. Table 4 shows the result of this experiment. Both systems perform better by deploying domain annotations, and TabVec achieves 5-13% higher F1 micro score. DWTC on the other side, performs much better when trained on the domain annotations. This shows that DWTC needs to be trained on the specific domain to perform well.

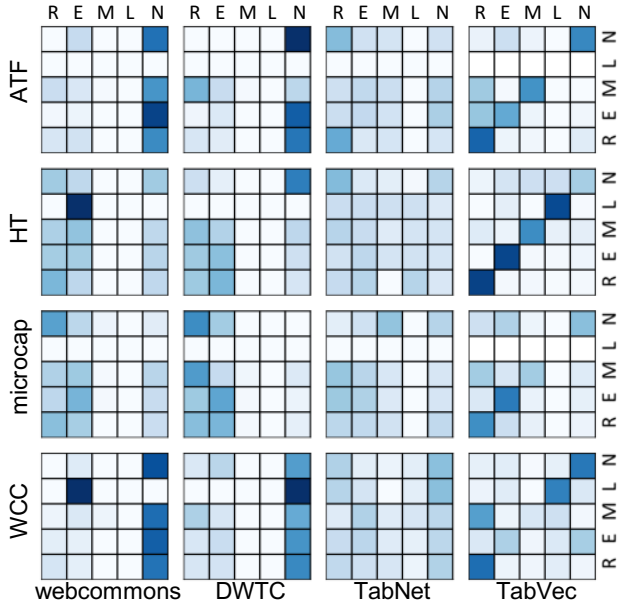

**Figure 9: classification confusion matrices for different domains and systems. Note that here *R, L, M, N*, and *E* stands for relational, list, matrix, non-data, and entity table types. Also, the X axis is the predicted label and the Y axis is the true label for all confusion matrices. Also note that for microcap domain we do not have a list table type.**

| | system | per class F1 score | | | | | F1-M |
|---|---|---|---|---|---|---|---|
| | | R | E | M | L | ND | |
| ATF | DWTC | **0.94** | **0.96** | **0.96** | - | **0.97** | **0.95** |
| | TabVec | 0.78 | 0.74 | 0.85 | - | 0.79 | 0.79 |
| HT | DWTC | **0.98** | **0.99** | **0.98** | 0.86 | **0.78** | **0.93** |
| | TabVec | 0.91 | 0.85 | 0.78 | 0.87 | 0.6 | 0.82 |
| mcap | DWTC | **0.83** | **0.95** | **0.74** | - | **0.92** | **0.87** |
| | TabVec | 0.62 | 0.68 | 0.48 | - | 0.58 | 0.61 |
| WCC | DWTC | **0.83** | **0.91** | **0.73** | 0.0 | **0.79** | **0.79** |
| | TabVec | 0.7 | 0.53 | 0.47 | **0.55** | 0.71 | 0.63 |

**Table 4: 10-fold cross validation results. Note that in this table *R, E, M, L, ND* stand for relational, entity, matrix, list, and non-data table types. Also, *F1-M* and *mcap* stands for F1-micro and microcap respectively.**

*4.2.4 Scalability discussion.* We can compare the scalability of our proposed system with the state of the art systems that we used as the baselines in this paper from two different aspects:

**(1) need for user intervention:** Although *TabVec* can benefit from manual annotations, and training data, this is not a necessary task. In other words, to get better performance for *webcommons*, *DWTC*, or *TabNet*, model training on the domain is necessary, which requires user annotations. Moreover, running *TabVec* is not preconditioned by inputing user annotations, and the user can provide annotations in an interactive way, and see the improvements. Other systems need to retrain their models with every new annotations provided. This makes *TabVec* more scalable compared to its rivals.

**(2) run-time:** Since all steps in our method can be parallelized, TabVec is easily scalable. Also, it needs finite memory (word vector space is the part that needs to fit in memory. Note that while there can be millions of pages in a domain, number of words usually will not exceed 100,000). We used Apache Spark (in standalone mode) to parallelize our computations. Table 5 shows the execution time for different steps of our method on WCC domain. We run this experiment on an 8-core desktop computer with 32GB of RAM. We observe that increasing the number of executors reduces the execution time (there is partitioning and job assigning overhead that is present in all columns). Note that for our system, as we go beyond 6 executors, the memory is overloaded (since there are shared resources, like word vector space, that is sent to all executors) which degrades the performance; so, having a cluster of nodes is expected to allow increasing the number of executors and improve the execution time. Also, please note that the first two steps are done to calculate the word vector space, which is done once per domain, and the table vector calculation is pretty fast (∼1.5ms per table in our experiment, since WCC has 20K tables in it).

| Execution time (s) | No. executors | | | | |
|---|---|---|---|---|---|
| | 1 | 2 | 4 | 6 | 8 |
| pre-process tables | 334 | 192 | 143 | 139 | 146 |
| train word vector space | 285 | 172 | 131 | 129 | 141 |
| calculate table vectors | 61 | 38 | 30 | 33 | 34 |
| Total | 680 | 402 | 304 | **301** | 321 |

**Table 5: Execution time in seconds for different steps of TabVec, on WCC domain.**

## 5 CONCLUSION

We proposed TabVec, an unsupervised method for constructing a table vector space from a corpus of web tables on specific domain, using different word context definitions for tables (surrounding text, cell text, header cells, and adjacent cells). Table vectors created by TabVec embed structure of web tables, and can be used to cluster tables according to their structural features. We showed that these table clusters may be deployed to classify tables by their type, i.e. the user labels each cluster with proper table type. In contrast to previous methods which rely on user annotations for training on each domain, or require an existing knowledge base, TabVec requires minimal user intervention, only to label a small number of clusters. Note that although we focused on web tables in this paper, our approach can be applied to non-html tables (CSV sheets, or PDF documents).

We evaluated our system on four real world domains (ATF, HT, microcap, and WCC), and compared its performance with three state of the art systems (webcommons, DWTC, and TabNet) for table type classification. Our experiments show that, when no domain knowledge is available, our system outperforms the baselines and *DWTC* comes second in all the domains. This shows that pretrained models provided for DWTC and webcommons systems do not perform well on our experimented domains. We also evaluated TabVec, TabNet and DWTC in a supervised classification setting. Our experiments revealed that TabNet cannot train a good model with our provided annotations. DWTC and TabVec leverage domain annotations to improve their performance. In case of having many annotated tables from a domain, DWTC (which is a feature based method) performs better than TabVec. Nevertheless, providing annotations is a cumbersome task and is not favorable in many applications.

In this paper we developed notions of cell vectors. In future work, we plan to explore the use of these cell vectors to extract knowledge from tables. Moreover, we observed that each context definitions for words in a table has its own power, and captures some types of tables better than others. Exploring a good way to combine these context definitions to construct a more robust word vector space is another topic for future work.

## ACKNOWLEDGMENTS

This research is supported by the Defense Advanced Research Projects Agency (DARPA) and the Air Force Research Laboratory (AFRL) under contract number FA8750-14-C-0240. The views and conclusions contained herein are those of the authors and should not be interpreted as necessarily representing the official policies or endorsements, either expressed or implied, of DARPA, AFRL, or the U.S. Government.